\documentclass[usenatbib]{mn2e}
\usepackage{apjfonts,amsfonts,amsmath,amssymb,bm,ctable,verbatim}

\newcommand{\acknowledgments}[1]{\begin{small}\section*{Acknowledgments}\end{small}{\noindent #1}\vspace{5pt}}

\title[Stellar Clustering \&\ Radiation Environments]{Most Stars (and Planets?) Are Born in Intense Radiation Fields}

\author[Lee \&\ Hopkins]{
\parbox[t]{\textwidth}{Eve J.~Lee$^{1,2}$ \&\ Philip F.~Hopkins$^1$} \vspace*{4pt} \\
$^1$ TAPIR, Mailcode 350-17, California Institute of Technology, Pasadena, CA 91125, USA\\
$^2$ Department of Physics and McGill Space Institute, McGill University, 3550 rue University, Montreal, QC, H3A 2T8, Canada
}

\date{}
\begin{document}
\maketitle

\begin{abstract}
Protostars and young stars are strongly spatially ``clustered'' or ``correlated'' within their natal giant molecular clouds (GMCs). We demonstrate that such clustering leads to the conclusion that the incident bolometric radiative flux upon a random young star/disc is enhanced (relative to volume-averaged fluxes) by a factor which increases with the total stellar mass of the complex. Because the Galactic cloud mass function is top-heavy,  the typical star in our Galaxy experienced a much stronger radiative environment than those forming in well-observed nearby (but relatively small) clouds, exceeding fluxes in the Orion Nebular Cluster by factors of $\gtrsim$30. Heating of the circumstellar disc around a median young star is dominated by this external radiation beyond $\sim 50\,$AU. And if discs are not well-shielded by ambient dust, external UV irradiation can dominate over the host star down to sub-AU scales. Another consequence of stellar clustering is an extremely broad Galaxy-wide distribution of incident flux (spanning $>10$ decades), with half the Galactic star formation in a substantial ``tail'' towards even more intense background radiation. We also show that the strength of external irradiation is amplified super-linearly in high-density environments such as the Galactic centre, starbursts, or high-redshift galaxies.
\end{abstract}


\section{Introduction}

\begin{figure}
    \centering
    \includegraphics[width=1.0\columnwidth]{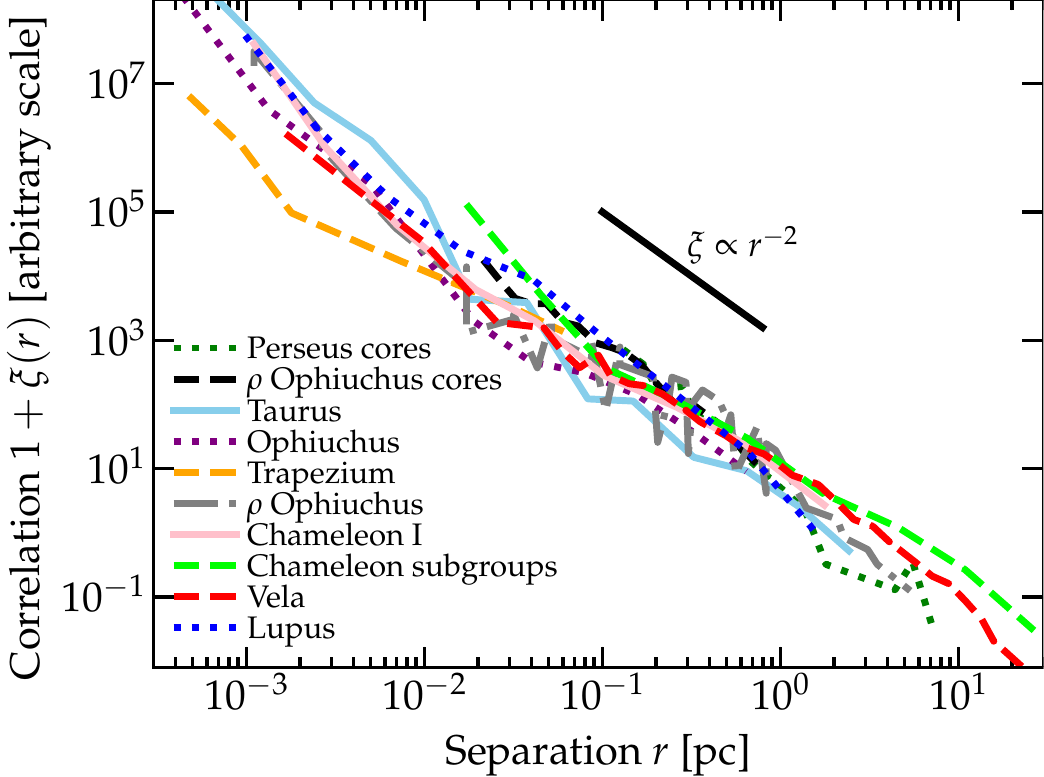}\\
    \hspace{0.2cm}\includegraphics[width=0.97\columnwidth]{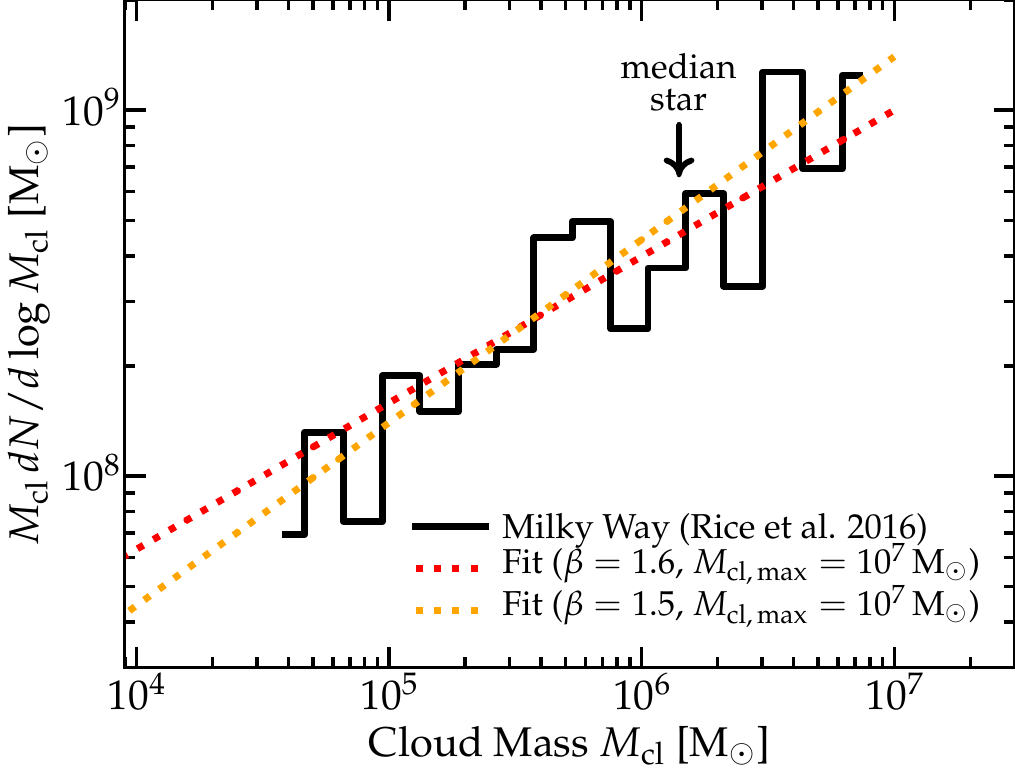}\\
    \vspace{-0.3cm}
    \caption{Observables motivating the arguments here. 
    {\em Top:} Observed clustering of young stars or protostellar cores \citep[compiled from][]{simon:1997.stellar.clustering,nakajima:1998.stellar.clustering,hartmann:2002.taurus.stellar.clustering,stanke:2006.core.mf.clustering,kraus:2008.stellar.clustering,enoch:2008.core.mf.clustering}: this is the correlation function $1+\xi(r)$, proportional to the mean density of stars at a given separation $r$ {\em from a star} (as opposed to a random point). We show the simple power law: $1+\xi(r) \propto r^{-2}$ which approximates the systems reasonably well over the range of interest. Note the normalization is arbitrary here, curves are normalized to overlap to clearly see they follow approximately the same power-law. 
    {\em Bottom:} Mass function of Milky Way (MW) molecular clouds, from \citet{rice:2016.gmc.mw.catalogue}. We compare fits with $dN/d M_{\rm cl} \propto M_{\rm cl}^{-\beta}$, with an upper limit observed at about $M_{\rm cl,\,max} \sim 10^{7}\,M_{\odot}$. We plot $M_{\rm cl}\,dN/d\log{M_{\rm cl}}$, i.e.\ the total mass per logarithmic interval in cloud mass (or size), to show the mass function is top-heavy ($\beta<2$). Arrow notes the median cloud mass in which a star would form (i.e.\ mass above which half the Galactic star formation is concentrated), assuming all have similar star formation efficiencies $\epsilon$. 
    \label{fig:demo.corrfn.mf}}
\end{figure}

It is a well-established observational fact that young stars, ``at birth,'' are strongly {\em statistically} clustered,\footnote{Note that we are careful to distinguish {\em clusters}, which are traditionally defined as specific objects, either self-gravitating (e.g.\ globulars) or above some surface density threshold, and {\em clustering}, which is a statistical property (``stars are more likely to form in regions that form other stars, as compared to randomly within the ISM'') that has nothing to do with bounded-ness.} meaning formally that they have a steep, non-zero auto-correlation function $\xi(r)$  -- in other words, stars are more likely to form near other stars than in isolation. Moreover the form of this correlation function appears to be near-universal for young stars across diverse star-forming regions (e.g.\ Taurus, Ophiuchus, Trapezium, Upper Sco, Chamaeleon, Orion, Lupus, Vela, 30 Dor, and more; see \citealt{simon:1997.stellar.clustering,nakajima:1998.stellar.clustering,hartmann:2002.taurus.stellar.clustering,hennekemper:2008.smc.stellar.clustering,kraus:2008.stellar.clustering}), and is consistent with the observed auto-correlation function of proto-stellar cores (\citealt{stanke:2006.core.mf.clustering,enoch:2008.core.mf.clustering};
Fig.~\ref{fig:demo.corrfn.mf} shows a collection of these data). Meanwhile, it is also well-established that the mass function of star-forming giant molecular clouds (GMCs) is ``top-heavy'' ($dN_{\rm cl}/dM_{\rm cl} \propto M_{\rm cl}^{-\beta}$ with $1.5<\beta<1.8$ up to a maximum ``cutoff'' mass) such that most of the star formation in the Galaxy (and in almost all other observed galaxies) occurs in the most massive clouds \citep{engargiola:2003.m33.gmc.catalogue,rosolowsky:gmc.mass.spectrum,audit:2010.gmc.massfunctions,2011ApJS..197...16W,2013ApJ...779...46H,2014A&A...567A.118D}. In the Milky Way (MW), the majority of all stars today are being formed in GMCs with masses $\gtrsim 10^{6}\,M_{\odot}$ 
(e.g., \citealt{williams:1997.gmc.prop,rice:2016.gmc.mw.catalogue,mamd17}; see Fig.~\ref{fig:demo.corrfn.mf}). In the star formation community, neither of these statements is controversial. Moreover, they are both predicted by essentially all theoretical models of star formation \citep[see e.g.][]{klessen:2000.cluster.formation,hansen:2012.lowmass.sf.radsims,hopkins:excursion.ism,hopkins:excursion.clustering,offner:2013.imf.review}, as \citet{guszejnov:universal.scalings} showed that both the qualitative form of $\xi(r)$ and $dN_{\rm cl}/dM_{\rm cl}$ are inevitable predictions of {\em any} model where star formation is the result of any scale-free hierarchical processes (including e.g.\ self-gravity, turbulence, magnetohydrodynamics). 

In this letter, we show that these observational facts lead immediately to the prediction that most stars in the Galaxy are born in environments with much higher stellar densities, and much stronger external radiation fields, than those present in commonly-observed {\em nearby} star-forming environments such as Taurus and even Orion \citep[e.g.][]{Fatuzzo08,Holden11,Adams12}. We illustrate the implications of the enhanced background radiation for circumstellar (protoplanetary) discs and their evolution. 
Some previous work have modeled the effects of external radiation in individual clusters (including globulars) for e.g.\ gas giant formation \citep[e.g.][]{Thompson13} or specifically on the truncation of protoplanetary discs \citep[e.g.][]{Winter18}. More generally, \citet{Winter20} have derived a distribution of disc truncation timescales in the solar neighbourhood and the central molecular zone due to external FUV radiation using a log-normal density distribution.
Our aim is different: we underscore the direct consequence of stellar clustering---not to be confused with the density distribution---on the background radiation fields experienced by typical stars forming in our Galaxy, across a variety of nebular environments.

\section{Observed Scalings and Consequences}

\subsection{Key Scalings}

How ``clustered'' stars are is encapsulated in $\xi(r)$ defined by
\begin{equation}
    1+\xi(r) \equiv \langle N(r)\rangle / \langle n\rangle\,dV
\end{equation}
where $\langle n \rangle$ is the mean number density of objects (stars), and $N(r)$ is the number of stars found at a radius $r$ from another star within a differential volume element $dV$. In other words, $\xi(r)$ describes the excess probability of finding a star around another star. We stress that $r$ is the distance between any random pair of stars, not the distance from e.g.\ the cloud centre.
In projection, $\xi$ is often quantified as the mean surface density profile of stars around a random star (just the line-of-sight-integral of $\xi$), $\Sigma(R) \propto  \int (1+\xi(r))\,dz$, and the observations (and theory) above consistently find $\Sigma(R) \propto R^{-1}$ ($1+\xi \propto r^{-2}$) for young stars and protostellar cores (with a steepening at small scales, $\lesssim 100\,$au, perhaps owing to the contribution of binaries; see, e.g. \citealt{ElBardry19} and \citealt{Moe19} for the evidence that binaries with separations $\lesssim 100\,$au likely emerged via disc instability rather than core fragmentation). 
As one looks at older and older stellar populations, this correlation function gradually ``flattens'' as most stars are not formed in {\em bound} clusters but loose associations, which disperse and mix until the old stars are more randomly distributed in the galaxy, but this process takes tens to hundreds of millions of years \citep[see e.g.][]{2019MNRAS.483.4707G}. 

So let us define $1+\xi \approx \xi_0 (r_{\rm cl}/r)^{\gamma} - \delta_{\xi}$ at $r\lesssim r_{\rm cl}$ where $r_{\rm cl}$ is some effective radius of a star-forming cloud (and $\delta_{\xi}$ is a small ``edge term'' that is only important at $r \gtrsim r_{\rm cl}$).\footnote{Cloud ``radii'' $r_{\rm cl}$ can sometimes be ambiguous. We could instead use the correlation length $r_{\rm corr}$ (scale where stellar density approaches the background mean) which can be un-ambiguously defined for any stellar field from just $\xi(r)$. But for a cloud with a well-defined outer profile (e.g.\ sharp or Gaussian ``edge''), this is trivially related (by an order-unity constant) to the usual $r_{\rm cl}$ defined from e.g.\ moments of the stellar distribution or density contours, so for clarity we work with the usual $r_{\rm cl}$. Likewise, we assume spherical clouds for simplicity; asymmetry will only introduce order-unity corrections so long as $r_{\rm cl}$ is defined as the geometric mean of the axes (a typical axis ratio of GMCs is $\sim2-3$, e.g. \citealt{2016A&A...593A...4M}).}
The normalization $\xi_0 \approx 1-\gamma/3$ (for the case of interest with $\gamma>1$) is determined by the definition of $\xi$ and finite number/mass of stars in a cloud. Within a spherical shell of volume $dV \equiv 4\pi r^2 dr$ (distance $r$) centred on any given star, ignoring edge effects ($r \ll r_{\rm cl}$), the average stellar mass density can be written as $\langle \rho_{\ast} (r) \rangle \approx \langle \rho_{\ast} \rangle_{\rm vol} (1+\xi(r)) \sim \langle \rho_{\ast} \rangle_{\rm vol}\,\xi(r)$ with $\langle \rho_{\ast} \rangle_{\rm vol} \equiv 3 M_{\ast}(r_{\rm cl}) / 4\pi r_{\rm cl}^3$. We can relate cloud mass and radius via surface density: $\Sigma_{\rm cl} \equiv M_{\rm cl}/\pi\,r_{\rm cl}^{2}$, and relate stellar and cloud mass via some star formation efficiency $\epsilon \equiv M_{\ast}(r_{\rm cl})/M_{\rm cl}$. 
This gives:
\begin{equation}
    \langle \rho_{\ast}(r) \rangle \sim \frac{3}{4}\left(1-\frac{\gamma}{3}\right)\epsilon \frac{\Sigma_{\rm cl}}{r_{\rm cl}}\left(\frac{r_{\rm cl}}{r}\right)^{\gamma}.
\end{equation}
with stellar mass enclosed $M_{\ast}(<r) \equiv \int_{0}^{r}\,\rho_{\ast}({\bf r}^{\prime})\,d^{3}{\bf r}^{\prime} \sim M_{\ast}(r_{\rm cl})\,(r/r_{\rm cl})^{3-\gamma}$. Substituting the observed $\gamma\approx 2$, 
\begin{align}
    \langle \rho_{\ast}(r) \rangle &\sim \frac{1}{4}\frac{\epsilon \Sigma_{\rm cl} r_{\rm cl}}{r^2} \sim \frac{\epsilon}{4r^2}\left(\frac{\Sigma_{\rm cl}M_{\rm cl}}{\pi}\right)^{1/2}.
    \label{eq:rhostar}
\end{align}
Note that on average, MW GMCs follow a well-defined scaling relation $M_{\rm cl} \propto r_{\rm cl}^2$, so have approximately constant (mass-independent)  $\Sigma_{\rm cl} \sim 100\,M_\odot\,{\rm pc}^{-2}$ \citep{bolatto:2008.gmc.properties,heyer:2009.gmc.trends.w.surface.density}; we will use this below.

\begin{figure*}
    \centering
    \includegraphics[width=0.48\textwidth]{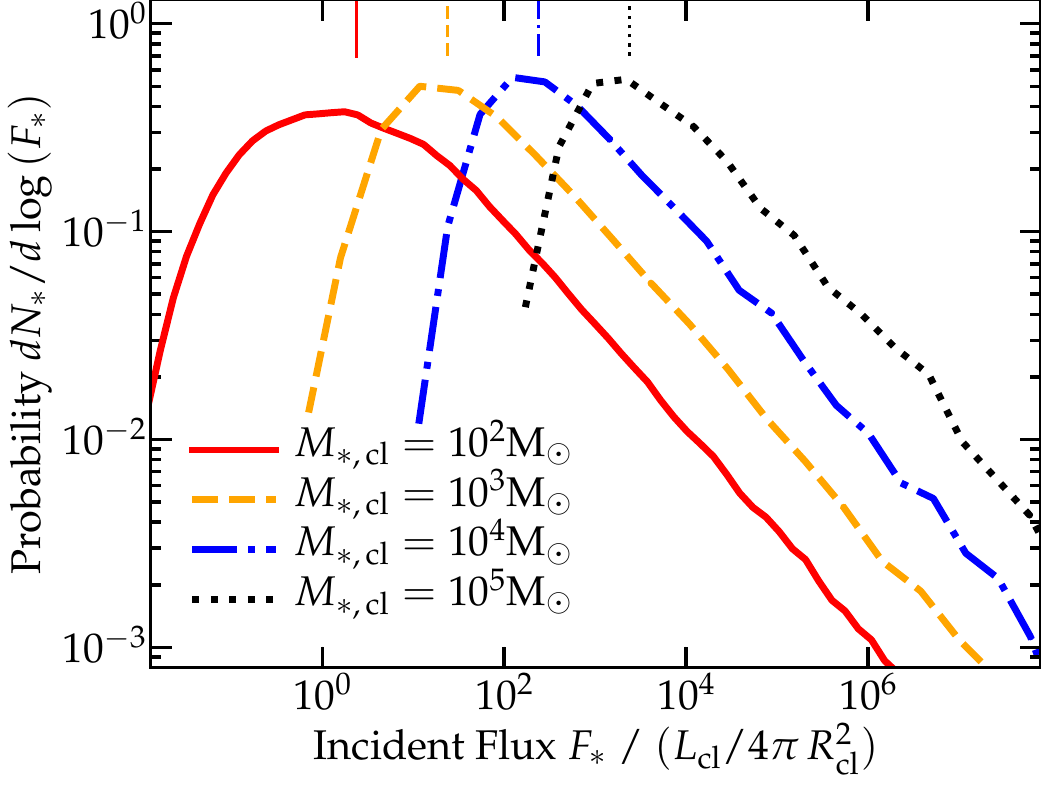} \includegraphics[width=0.48\textwidth]{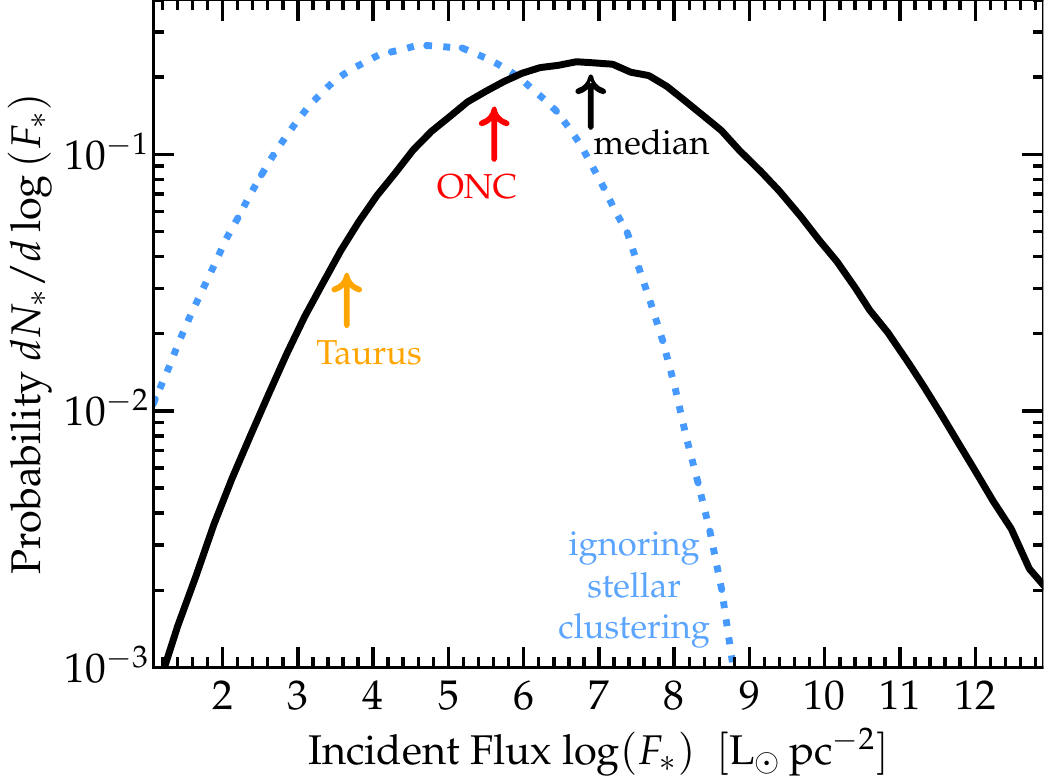}
    \vspace{-0.3cm}
    \caption{Results of stellar clustering. 
    {\em Left:} Distribution of incident fluxes on the location of a random star. 
    For each realization (corresponding to each random, reference star), stellar masses are drawn from the initial mass function \citep{kroupa:2001.imf.var}, summed up to match the given total stellar mass $M_{\rm \ast,\,cl}$. Relative radial positions of all member stars from the reference star are drawn so that their number density follows the observed correlation function $1+\xi(r) \propto r^{-2}$ (i.e., the enclosed number distribution should follow $r$). We adopt the Starburst99 model \citep{starburst99} to convert the masses to zero-age main-sequence bolometric luminosities. Ignoring attenuation, we plot the distribution of incident bolometric fluxes on each reference star, normalized by the mean volume-averaged flux or luminosity per unit area at a random point (not a random star), 
    $\sim L_{\rm cl} / 4\pi R_{\rm cl}^2 \equiv \langle L_{\ast}/m_{\ast} \rangle  M_{{\ast},\rm cl} / 4\pi\,R_{\rm cl}^{2} = \langle L_{\ast}/m_{\ast} \rangle \epsilon \Sigma_{\rm cl}/4$---this normalization factor stays constant irrespective of cluster/complex mass $M_{\ast,\,{\rm cl}}$. We note that the median incident flux is larger in more massive systems. Vertical lines show the predicted median from Eq.~\ref{eqn:favg}, which agrees extremely well with the full Monte Carlo calculation shown. Note also the broad tail of the distribution, with $d N_{\ast} / d\log{(F_{\ast})} \propto F_{\ast}^{-1/2}$ at high-$F_{\ast}$.
    {\em Right:} Distributions at {\em left}, convolved over the observed Milky Way cloud mass function and distribution of $L_{\ast}/4\pi\,R_{\rm cl}^{2}$ (calculated by taking the distribution of cloud gas mass from \citealt{rice:2016.gmc.mw.catalogue}, and cloud surface densities $\Sigma_{\rm cl}$ and the star formation efficiency $\epsilon$ from \citealt{lee:mw.cloud.dynamical.sfe}; replacing the cloud gas mass distribution with those in \citealt{lee:mw.cloud.dynamical.sfe} makes negligible difference), weighted by the number of stars formed with this incident $F_{\ast}$. 
    Taking into account stellar clustering systematically shifts the entire flux distribution to high values while extending the high flux tail.
    We label the median value $\sim 10^{6.8}\,{\rm L_{\odot}\,pc^{-2}}$, and the best estimate of $\langle F_{\ast} \rangle$ for the ONC and Taurus \citep[e.g.][and references therein]{Kryukova12,Marks12}, using the same conversions as before. The ONC (Taurus) lies a factor $\sim 30$ ($\sim1000$) below the median, because their masses are diminutive on a Galactic scale.
    Most stars are produced in the most massive clouds, and therefore subject to higher $F_{\ast}$.
    \label{fig:flux.distrib}}
\end{figure*}

It is now straightforward to estimate the average incident {\em bolometric} radiation flux, $\langle F \rangle = \int (1/4\pi\,r^{2})\,(L_{\ast}/m_{\ast})\,dm_{\ast} = \int \langle L_{\ast}/m_{\ast} \rangle \,(\rho_{\ast}(r)/4\pi\,r^{2})\,d^{3}{\bf r}$, where $\langle L_{\ast}/m_{\ast} \rangle \approx 1300\,L_{\odot}/M_{\odot}$ is itself averaged over the stellar initial mass function (IMF) at each point for a zero-age main sequence population \citep{starburst99}, integrating from some minimum $r_{\rm min}$ to $r_{\rm cl}$. But this diverges as $\langle F \rangle \approx (r_{\rm cl}/r_{\rm min})\, \langle L_{\ast}/m_{\ast} \rangle\,M_{\rm cl}/(4\pi\,r_{\rm cl}^{2}) \propto 1/r_{\rm min}$: indicating that the contributions are dominated by the closest stars, for which we need to account for finite sampling effects. We can approximate the {\em median} results of a detailed Monte Carlo IMF-sampling calculation (shown in Fig.~\ref{fig:flux.distrib}) by simply estimating $r_{\rm min}$ as the mean distance enclosing some critical stellar mass sufficient for an order-unity probability of a single intermediate-mass star, $\int_{0}^{r_{\rm min}}\,\rho_{\ast}(r)\,d^{3}{\bf r} = \langle m_{\ast}^{\rm eff} \rangle \sim 30\,M_{\odot}$, giving: 
\begin{align}
    \label{eq:ravg}
    \langle r_{\rm min} \rangle &\sim \frac{\langle m_{\ast}^{\rm eff} \rangle}{\epsilon} \left(\frac{1}{\pi\Sigma_{\rm cl}M_{\rm cl}}\right)^{1/2} \\ 
    \label{eqn:favg} \langle F_{\ast} \rangle &\sim \frac{\langle L_{\ast}/m_{\ast} \rangle}{4} \epsilon^2 \frac{\Sigma_{\rm cl}M_{\rm cl}}{\langle m_{\ast}^{\rm eff} \rangle} 
   \sim 10^{6}\,\epsilon_{0.05}^{2}\,\langle \Sigma_{\rm cl,100} \rangle\,M_{\rm cl,6}\,\frac{L_{\odot}}{\rm pc^{2}}
\end{align}
where $\langle \Sigma_{\rm cl,\,100} \rangle \equiv \langle \Sigma_{\rm cl} \rangle/100\,M_{\odot}\,{\rm pc^{-2}} \sim M_{\rm cl,\,6}/R_{\rm cl,\,50}^{2}$, $M_{\rm cl,\,6}\equiv M_{\rm cl}/10^{6}\,M_{\odot}$, $R_{\rm cl,\,50}\sim R_{\rm cl}/50\,{\rm pc}$, and $\epsilon_{0.05}\equiv\epsilon/0.05$.
Finally, averaging $\langle F_{\ast} \rangle$ over the cloud mass function $dN/d M_{\rm cl} \propto M_{\rm cl}^{-\beta}$ up to a maximum cloud mass $M_{\rm cl,\,max}$ for observed $1.5 < \beta < 1.8$ (weighting by number of stars per cloud, and using the fact that $\epsilon\sim 0.05$ shows no systematic dependence on $M_{\rm cl}$ in the MW; \citealt{lee:mw.cloud.dynamical.sfe,grudic:sfe.gmcs.vs.obs,2019arXiv190511827W}) gives:
\begin{align}
\label{eqn:flux.mf.avg} \langle \langle F_{\ast} \rangle \rangle &\sim 0.3\,\langle F_{\ast}(M_{\rm cl}=M_{\rm cl,\,max}) \rangle 
\sim \frac{0.07\,\langle L_{\ast}/{m_{\ast}}\rangle}{\langle m_{\ast}^{\rm eff} \rangle} \,{\epsilon^{2}\,\Sigma_{\rm cl}\,M_{\rm cl,\,max}}
\end{align}

\subsection{Implications: Stars in Different Clouds}

We can re-write Eq.~\ref{eqn:favg} as $\langle F_{\ast} \rangle \sim (\epsilon\,M_{\rm cl} / 4\,\langle m_{\ast}^{\rm eff} \rangle) \, (\langle L_{\ast}/{m_{\ast}}\rangle \,\epsilon\,\Sigma_{\rm cl}) = (L_{\rm cl}/4\pi\,R_{\rm cl}^{2})\,(M_{\ast,\,\rm cl}/30\,M_{\odot})$, where $(L_{\rm cl}/4\pi\,R_{\rm cl}^{2}) = \langle L_{\ast}/m_{\ast} \rangle\,\epsilon\,\Sigma_{\rm cl}/4$ is the {\em mean} luminosity per unit area averaged over the entire volume of the cloud. In other words, the flux density {\em at the locations of individual stars} is enhanced (because their locations are correlated) by the cloud-mass-dependent factor $M_{\ast,\,\rm cl}/\langle m_{\ast}^{\rm eff} \rangle \propto \epsilon\,M_{\rm cl}$. 
Fig.~\ref{fig:flux.distrib} illustrates this enhancement. We observe the systematic shift of the flux distribution for more massive complexes; this $F_{\ast}-M_{\rm \ast,\,cl}$ dependence is lost when stellar clustering is non-existent or weak \citep[e.g. see][their equation 1]{Fatuzzo08}. 
Another effect of clustering is the elongation of the high flux tail (see Fig.~\ref{fig:flux.distrib}, right panel). Such elongation is caused by both the mass-dependent flux enhancement as well as the heightened sensitivity to the sampling of the stellar IMF. If stars were randomly and uniformly placed in any given system (no clustering, $\gamma=0$), then the total incident flux on a random star from other stars at a distance $r$ would scale as $\propto (r/R_{\rm cl})$ and therefore $F$ would be dominated by the collective light of all the stars (i.e., stars enclosed within large distances $\sim R_{\rm cl}$). This acts to reduce star-to-star variation in the incident flux (both from IMF and stellar position sampling). If, on the other hand, stars are clustered (with $\gamma > 1$), then the total incident flux on a random star scales as $(R_{\rm cl}/r)^{\gamma-1}$ and so becomes sensitive to the positions and the masses of the closest neighbours. This increase in star-to-star variation is reflected in the larger high flux tail in both intra- and inter-cloud incident flux distributions illustrated in Fig.~\ref{fig:flux.distrib}.

This is mathematically obvious, but leads to a crucial non-intuitive result. As noted above, $\epsilon$ and $\Sigma_{\rm cl}$ are roughly independent of cloud mass, so the mean {\em cloud-volume-averaged} stellar surface density or flux $\sim \epsilon\,\Sigma_{\rm cl} \propto L_{\rm cl}/4\pi\,R_{\rm cl}^{2}$ is independent of cloud mass, and the {\em three-dimensional} cloud gas or stellar density $\langle \rho_{\ast}\rangle_{\rm vol} \propto M_{\rm cl}/R_{\rm cl}^{3} \propto \Sigma_{\rm cl} / R_{\rm cl} \propto R_{\rm cl}^{-1}$ actually {\em decreases} in larger/more-massive clouds. But, because of clustering, the mean density of stars at fixed distance ($\langle \rho_{\ast}(r) \rangle$) or incident radiative flux ($\langle F_{\ast} \rangle$) {\em around a star} actually {\em increases} in more massive clouds/complexes/clusters. 
In short, a generic result of stellar clustering is that stars forming in more massive complexes experience systematically different radiative environments.\footnote{For an arbitrary $\gamma$, $\langle F_{\ast} \rangle \sim  (L_{\rm cl}/4\pi\,R_{\rm cl}^{2})\,q\,(M_{\ast,\,\rm cl}/\langle m_{\ast}^{\rm eff} \rangle)^{q}$ with $q\equiv (\gamma-1)/(3-\gamma) \approx \gamma_{\rm 2D}/(2-\gamma_{\rm 2D})$, where $\gamma_{\rm 2D}$ is given by the two-dimensional (projected/observed) correlation function $\omega(R) \propto R^{-\gamma_{\rm 2D}}$. So any Universe where the stellar correlation function (in projection) rises at small $R$ leads to $\langle F_{\ast} \rangle$ increasing with $M_{\ast,\,\rm cl}\equiv M_{\ast}(r_{\rm cl}) \equiv \epsilon\,M_{\rm cl}$.}

In Eq.~\ref{eqn:flux.mf.avg}, we see an immediate consequence of this: the median $\langle F_{\ast} \rangle$ increases linearly with total cloud mass, and because the GMC mass function is top-heavy, the median star in the MW experiences a flux comparable to that in the most massive clouds: $\langle \langle F_{\ast} \rangle \rangle \sim 0.3\, \langle F_{\ast}  (M_{\rm cl}=M_{\rm cl,\,max}) \rangle$.  In contrast, the best-observed clouds (particularly for e.g.\ disc studies) tend to be those nearest to Earth, which of course means they are among the least massive (most numerous, but negligible for the total Galactic star formation rate) clouds, implying that the typical radiative environments are {\em much} weaker in these clouds. For example: Taurus, Perseus, Serpens, Ophiuchus, and Lupus (with $M_{\rm cl} \sim 10^{4}\,M_{\odot}$) have $\langle F_{\ast} \rangle$ a factor $\gtrsim 100-300$ weaker than the median MW star (assuming $M_{\rm max,\,6} \sim 3-10$), while even ``extreme'' or ``massive'' nearby regions such as Orion-A/B (or Monoceros, $\lambda$-Ori, Cepheus, Chameleon, Auriga) with $M_{\rm cl} \sim 10^{5}\,M_{\odot}$ have factor $\sim 30$ lower $\langle F_{\ast} \rangle$. In short, the median MW star today is forming in an environment with external radiation $\sim 30$ times stronger than the ``strongly irradiated'' Orion environments!

\subsection{Dependence on Galactic Environment}

We stress that we have adopted scalings appropriate for {\em field} star formation in normal, low-star-formation-efficiency clouds and loose associations -- we are not assuming all stars form in dense ``clusters'' for example. In fact we totally neglect especially dense systems like globulars, where of course $\langle F_{\ast} \rangle$ could be even higher 
\citep[see, e.g.][their Figure 6]{Thompson13}
but these contribute a small fraction of stars. 

It is worth noting that $\langle F_{\ast} \rangle \propto \epsilon^{2}\,\Sigma_{\rm cl}\,M_{\rm cl}$, and there is scatter and systematic variation in $\Sigma_{\rm cl}$ and $\epsilon$. In observations, numerical simulations, and simple analytic feedback-regulated models \citep{fall:2010.sf.eff.vs.surfacedensity,grudic:sfe.cluster.form.surface.density,grudic:mond.accel.scale.from.stellar.fb,2019MNRAS.487..364L,2019arXiv190511827W}, $\epsilon \sim (1 + 2000\,{\rm M_{\odot}\,pc^{-2}}/\Sigma_{\rm cl})^{-1}$ is generally found ($\epsilon \propto \Sigma_{\rm cl}$ up to a saturation level at $\epsilon \sim 1$), which gives $\langle F_{\ast} \rangle \propto \Sigma_{\rm cl}^{3}\,M_{\rm cl} \propto R_{\rm cl}^{2}\,\Sigma_{\rm cl}^{4}$, a highly super-linear dependence. In contrast, the naive assumption that $ F_{\ast}  \sim L_{\rm cl} / R_{\rm cl}^{2}$ (ignoring the fact that star formation is correlated, so assuming stars are uniformly distributed in volume in a cloud) would give $\langle F_{\ast} \rangle \propto \epsilon\,\Sigma_{\rm cl} \propto \Sigma_{\rm cl}^{2}$. So for clouds with even modestly-enhanced $\Sigma_{\rm cl}$, $\langle F_{\ast} \rangle$ can be much higher. 

This is especially important in the Galactic nucleus and in starburst or high-redshift galaxies, where $\Sigma_{\rm cl}$ can reach $\sim 10^{3}-10^{4}\,{\rm M_{\odot}\,pc^{-2}}$ \citep[see][]{kennicutt98,daddi:2010.ks.law.highz,genzel:2010.ks.law},
implying that $\langle F_{\ast} \rangle$ can be larger by factors $\gtrsim 10^{4}$ compared to nearby clouds, or (equivalently) that the distance to the nearest massive stars can be factors $\gtrsim  100$ smaller (reaching as small as $\lesssim 1\,$au, in the densest environments).

\subsection{Effect on Circumstellar Discs}

For a young star+disc bathed in an external bolometric radiation flux $F_{\ast}$, this flux will dominate over the incident flux from the central star outside a radius $r_{\ast} \gtrsim (L_{\ast}/4\pi\,F_{\ast})^{1/2} \sim 20\,{\rm AU}\,(L_{\ast}/L_{\odot})^{1/2}\,F_{\ast,7}^{-1/2}$ (where $F_{\ast, 7} \equiv F_{\ast} / 10^{7}\,L_{\odot}\,{\rm pc^{-2}}$). Taking $F_{\ast}\sim \langle F_{\ast} \rangle$ we obtain:
\begin{align}
    \langle r_{\ast} \rangle & \sim 50\,{\rm AU}\,\epsilon_{0.05}^{-1}\,\left( \frac{(L_{\ast}/L_{\odot})}{\langle \Sigma_{\rm cl,100} \rangle\,M_{\rm cl,6}} \right)^{1/2} \sim 50\,{\rm AU}\frac{(L_{\ast}/L_{\odot})^{1/2}}{\epsilon_{0.05} \langle \Sigma_{\rm cl,100}\rangle R_{\rm cl,60}} 
\end{align}
where $R_{\rm cl,60} \equiv R_{\rm cl}/60\,{\rm pc}$.
If the disc is optically-thick and in equilibrium, we can equate this to the surface flux $\sim \sigma\,T_{\rm eff}^{4}$ to show that this incident $F_{\ast}$ sets a ``floor'' or minimum to the surface effective temperature $T_{\rm eff}$, outside of $r\gtrsim r_{\ast}$:
\begin{align}
T_{\rm eff}^{\rm ext}  &\sim \left( \frac{F_{\ast}}{\sigma} \right)^{1/4} \sim 90\,F_{\ast,7}^{1/4}\,{\rm K} \\ 
\nonumber \langle T_{\rm eff}^{\rm ext} \rangle &\sim 60\,{\rm K}\,({\epsilon_{0.05} \langle \Sigma_{\rm cl,100}\rangle R_{\rm cl,60}})^{1/2},
\end{align}
which is greater than the condensation temperatures of common species such as methane, carbon monoxide, nitrogen ($N_2$) and hydrogen sulfide \citep[see][their Table 2 and references therein]{Zhang15}. It may be that in typical star forming environments in the Galaxy, many of these ice lines are non-existent and therefore do not play a role in planet formation.

For typical parameters in the outer disc, heating by external radiation can easily dominate over heating by disc accretion, and ensure the Toomre $Q\gg 1$ at $r \gtrsim r_{\ast}$ (where, under illumination by the central star alone, $Q$ would be decreasing) for essentially any disc with $M_{\rm disk} \lesssim 0.1\,M_{\ast}$. 
We note that the implication of stabilization of the disc by external heating reflects and actually enhances the findings of \citet{Thompson13} as we find higher $T^{\rm ext}_{\rm eff}$ and stronger dependence on $\Sigma_{\rm cl}$, from stellar clustering.

In extreme starburst environments ($\langle \Sigma_{\rm cl,100}\rangle \sim 10-100$) or for the ``tail'' of the distribution of high $F_{\ast}$,  $\langle T_{\rm eff}^{\rm ext} \rangle$ can exceed $\sim 300-1000$\,K (with $\langle r_{\ast} \rangle \lesssim $\,AU). Under these circumstances, the ambient radiation could heat gas to the point where protostellar accretion and/or disc formation are strongly suppressed, even modifying the stellar IMF itself \citep[see e.g.][]{guszejnov.2015:feedback.imf.invariance}. Such intense background radiation would effectively wipe out all ice lines in protoplanetary discs, potentially modifying the chemical make-up of the atmospheres of planets they may spawn. Dust grains would need to rapidly grow beyond a few $\mu$m's before they are radially drifted in to survive and eventually build up into planetary cores.

{\em If} we ignore attenuation, the effect of nearby stars is much more extreme in hard/ionizing bands (FUV/EUV/soft X-rays). A single typical O-star produces a tremendous amount of FUV/EUV/X-ray radiation, with ionizing $L_{\ast}^{\rm EUV} \gtrsim 10^{38}\,{\rm erg\,s^{-1}}$, while even a relatively ``active'' M-dwarf produces just $L_{\ast}^{\rm FUV} \sim 10^{26}\,{\rm erg\,s^{-1}}$ \citep{2013ApJ...763..149F,2018ApJS..239...16F}. From our full Monte Carlo calculation, the median distance to the nearest O-star is simply given by the distance $\langle r_{\rm min} \rangle$ enclosing $\langle m_{\ast}^{\rm eff} \rangle \sim 100\,M_{\odot}$ (as this gives an order-unity probability of one O-star from IMF sampling): putting these together, we obtain the result that around an M or FGK dwarf, the FUV/EUV/soft X-ray flux from the nearest O-star will dominate over that produced by the central dwarf star outside a radius  
$\langle r_{\ast}^{\rm EUV} \rangle \sim 0.04\,{\rm AU}\,(L_{\rm M,\,26}/L_{\rm O,\,38})^{1/2} / ({\epsilon_{0.05} \langle \Sigma_{\rm cl,100}\rangle R_{\rm cl,60}})$ where $L_{\rm M,\,26}$ is the UV luminosity of M dwarfs in units of $10^{26}\,{\rm ergs}\,{\rm s}^{-1}$, and $L_{\rm O,\,38}$ is the UV luminosity of O stars in units of $10^{38}\,{\rm ergs}\,{\rm s}^{-1}$. 
This is roughly the distance beyond which we find most planets \citep[e.g.][]{Dressing15,Lee17,Petigura18}, implying that in massive associations or complexes that contain at least one O star---which is where the majority of Galactic star formation occurs---the background EUV radiation field dominates over the central star across the entire planet-forming region (assuming there is minimal attenuation). 

The gravitational (tidal) effects of stellar clustering are generally weaker (as they drop off as $\sim 1/r^{3}$, and involve mass which scales less steeply than luminosity). For massive stars, nearly all are observed to be in binaries, with those binaries strongly biased towards equal-mass, so the effects of the nearest stars outside the binary (from larger-scale clustering) will be much weaker than the effect of the binary companion on the disc. However for e.g.\ an M-dwarf which may not have a close companion, the external tidal effect could be important. The $\sim m/r^{3}$ scaling of tidal forces means this will be dominated by the nearest companion of broadly-similar mass: for an M-dwarf, this is at a distance $\langle r_{\rm min} \rangle$ enclosing $\langle m_{\ast}^{\rm eff} \rangle \sim 0.3\,M_{\odot}$, which is just $\sim 70\,{\rm AU}/({\epsilon_{0.05} \langle \Sigma_{\rm cl,100}\rangle R_{\rm cl,50}})$. So tidal effects will generally only be important at larger radii compared to radiative effects, but these radii can still be relatively small.

\subsection{Obscuration and Time-Dependence}

A major uncertainty in all the above is time-dependence. We have taken all our scalings from a ``snapshot'' in time for young stars and/or protostars, as observed. But this should not be taken to imply any ``steady-state''. 
In fact, it is likely that the observed star forming regions are sampled from a variety of evolutionary stages with explicitly time-varying star formation efficiencies $\epsilon$ \citep{lee:mw.cloud.dynamical.sfe}. If all clouds were co-eval, at the zero-th order, the incident flux distribution would simply gradually shift to higher fluxes over time (assuming the correlation function and the stellar IMF are time-invariant).  
Evolution of star-forming clouds is a more complex process:
continuing star formation means new stars are formed; HII regions and wind bubbles around massive stars expand (potentially engulfing other stars/cores), overlap, and can disrupt the GMC as a whole; massive stars reach the end of their lives and explode; close passages (from un-bound stars) can occur; and stars drift and/or undergo hierarchical dynamical relaxation on a few crossing times (modifying the correlation function). At the first order, dynamical relaxation would lead to stellar mass segregation. On the assumption that the stars remain ``clustered'', mass segregation can broaden the distribution of incident flux on random stars within a star forming complex. Near the centre of the complex, the median background radiation enhances even more as the closest neighbours---which carries the lion's share of total incident flux on a random star---are more massive and luminous. By the same logic, however, the median radiation flux can drop considerably near the edges of the complex as the nearest neighbours are wimpy M dwarfs.
We note however that properly accounting for all of these (highly non-linear) effects requires detailed numerical simulations, and is well outside the scope of our calculation here.

For these reasons, the effects of obscuration are also difficult to estimate. We focus primarily on the bolometric flux (and tidal forces) above because these are (relatively) insensitive to the intervening dust and gas between stars. But the incident FUV/EUV/X-ray flux from e.g.\ nearby massive stars, which we noted above could (in principle, if un-obscured) play a major role in heating or evaporating discs at even quite small radii ($\ll 1\,$AU) around their parent stars, is of course strongly sensitive to the amount of ``shielding'' by both intervening dust and gas and the remaining dust and gas in the ``envelopes'' around both the emitting and receiving stars \citep[see, e.g.,][]{Gaches18}. But that dust and gas is, in turn, strongly sensitive to all of the above dynamical processes.

\section{Conclusions}

Stars are born clustered, with the correlation function $\xi \propto r^{-2}$. In the Milky Way Galaxy, most stars form in the most massive clouds, with masses $\gtrsim 10^6 M_\odot$. Through analytic and Monte Carlo calculations, we demonstrated that these two facts conspire to produce intense radiation environments in typical star-forming clouds and young clusters/associations, with the median incident flux on a star forming today in the Milky Way at least a factor 30 times higher than the fluxes in the ONC.

We explored two important consequences of stellar clustering. First, more massive star-forming regions experience more intense radiation (at birth). In particular, we showed that the mean incident flux on a random star within a given cloud or cluster is enhanced (relative to the flux averaged over the entire cloud volume) by a factor proportional to the total stellar mass of the complex: $\langle F_{\ast} \rangle \sim (L_{\rm cl}/4\pi R_{\rm cl}^2)\, (M_{\rm \ast,\,cl}/\langle m^{\rm eff}_{\ast}\rangle)$. Second, the distribution of incident flux develops a broad ``tail'' towards high fluxes, because the radiation seen by each star is dominated by the closest massive neighbours, rather than the sum of the entire cloud population. The diversity of separations and masses of neighbouring stars is therefore reflected in a broader Galactic distribution of incident fluxes.

Our calculations suggest that the typical star forming environments in the Milky Way Galaxy, and therefore, planet forming environments are likely bathed in radiation much more intense than we are accustomed to in our Solar neighbourhood and in nearby well-studied regions such as Taurus and even the ONC. The exact consequences of this for e.g.\ the structure and evolution of protostellar/protoplanetary discs, and therefore the likelihood and timing of planet formation, is an important question for future work. Exploring this in detail will require more careful calculations of the time-dependence of these processes as well as role of dust and gas in the stellar/planet nursery (potentially shielding short-wavelength radiation).

\acknowledgments{We thank Ruobing Dong and Stella Offner for useful feedback, Andrew Winter for interesting discussions, and the anonymous referee for their encouraging report. EJL was supported by the Sherman Fairchild Fellowship at Caltech. Support for PFH was provided by NSF Collaborative Research Grants 1715847 \&\ 1911233, NSF CAREER grant 1455342, NASA grants 80NSSC18K0562, JPL 1589742.}

\bibliography{ms_extracted}

\begin{thebibliography}{}
\makeatletter
\relax
\def\mn@urlcharsother{\let\do\@makeother \do\$\do\&\do\#\do\^\do\_\do\%\do\~}
\def\mn@doi{\begingroup\mn@urlcharsother \@ifnextchar [ {\mn@doi@}
  {\mn@doi@[]}}
\def\mn@doi@[#1]#2{\def\@tempa{#1}\ifx\@tempa\@empty \href
  {http://dx.doi.org/#2} {doi:#2}\else \href {http://dx.doi.org/#2} {#1}\fi
  \endgroup}
\def\mn@eprint#1#2{\mn@eprint@#1:#2::\@nil}
\def\mn@eprint@arXiv#1{\href {http://arxiv.org/abs/#1} {{\tt arXiv:#1}}}
\def\mn@eprint@dblp#1{\href {http://dblp.uni-trier.de/rec/bibtex/#1.xml}
  {dblp:#1}}
\def\mn@eprint@#1:#2:#3:#4\@nil{\def\@tempa {#1}\def\@tempb {#2}\def\@tempc
  {#3}\ifx \@tempc \@empty \let \@tempc \@tempb \let \@tempb \@tempa \fi \ifx
  \@tempb \@empty \def\@tempb {arXiv}\fi \@ifundefined
  {mn@eprint@\@tempb}{\@tempb:\@tempc}{\expandafter \expandafter \csname
  mn@eprint@\@tempb\endcsname \expandafter{\@tempc}}}

\bibitem[\protect\citeauthoryear{{Adams}, {Fatuzzo}  \& {Holden}}{{Adams}
  et~al.}{2012}]{Adams12}
{Adams} F.~C.,  {Fatuzzo} M.,   {Holden} L.,  2012, \mn@doi [\pasp]
  {10.1086/668084}, \href
  {https://ui.adsabs.harvard.edu/abs/2012PASP..124..913A} {124, 913}

\bibitem[\protect\citeauthoryear{{Audit} \& {Hennebelle}}{{Audit} \&
  {Hennebelle}}{2010}]{audit:2010.gmc.massfunctions}
{Audit} E.,  {Hennebelle} P.,  2010, \mn@doi [\aap]
  {10.1051/0004-6361/200912695}, \href
  {http://adsabs.harvard.edu/abs/2010A%26A...511A..76A} {511, A76+}

\bibitem[\protect\citeauthoryear{{Bolatto}, {Leroy}, {Rosolowsky}, {Walter}  \&
  {Blitz}}{{Bolatto} et~al.}{2008}]{bolatto:2008.gmc.properties}
{Bolatto} A.~D.,  {Leroy} A.~K.,  {Rosolowsky} E.,  {Walter} F.,   {Blitz} L.,
  2008, \mn@doi [\apj] {10.1086/591513}, \href
  {http://adsabs.harvard.edu/abs/2008ApJ...686..948B} {686, 948}

\bibitem[\protect\citeauthoryear{{Daddi} et~al.}{{Daddi}
  et~al.}{2010}]{daddi:2010.ks.law.highz}
{Daddi} E.,  et~al., 2010, \mn@doi [\apjl] {10.1088/2041-8205/714/1/L118},
  \href {http://adsabs.harvard.edu/abs/2010ApJ...714L.118D} {714, L118}

\bibitem[\protect\citeauthoryear{{Dressing} \& {Charbonneau}}{{Dressing} \&
  {Charbonneau}}{2015}]{Dressing15}
{Dressing} C.~D.,  {Charbonneau} D.,  2015, \mn@doi [\apj]
  {10.1088/0004-637X/807/1/45}, \href
  {https://ui.adsabs.harvard.edu/abs/2015ApJ...807...45D} {807, 45}

\bibitem[\protect\citeauthoryear{{Druard} et~al.,}{{Druard}
  et~al.}{2014}]{2014A&A...567A.118D}
{Druard} C.,  et~al., 2014, \mn@doi [\aap] {10.1051/0004-6361/201423682}, \href
  {https://ui.adsabs.harvard.edu/abs/2014A&A...567A.118D} {567, A118}

\bibitem[\protect\citeauthoryear{{El-Badry} \& {Rix}}{{El-Badry} \&
  {Rix}}{2019}]{ElBardry19}
{El-Badry} K.,  {Rix} H.-W.,  2019, \mn@doi [\mnras] {10.1093/mnrasl/sly206},
  \href {https://ui.adsabs.harvard.edu/abs/2019MNRAS.482L.139E} {482, L139}

\bibitem[\protect\citeauthoryear{{Engargiola}, {Plambeck}, {Rosolowsky}  \&
  {Blitz}}{{Engargiola} et~al.}{2003}]{engargiola:2003.m33.gmc.catalogue}
{Engargiola} G.,  {Plambeck} R.~L.,  {Rosolowsky} E.,   {Blitz} L.,  2003,
  \mn@doi [\apjs] {10.1086/379165}, \href
  {http://adsabs.harvard.edu/abs/2003ApJS..149..343E} {149, 343}

\bibitem[\protect\citeauthoryear{{Enoch}, {Evans}, {Sargent}, {Glenn},
  {Rosolowsky}  \& {Myers}}{{Enoch}
  et~al.}{2008}]{enoch:2008.core.mf.clustering}
{Enoch} M.~L.,  {Evans} II N.~J.,  {Sargent} A.~I.,  {Glenn} J.,  {Rosolowsky}
  E.,   {Myers} P.,  2008, \mn@doi [\apj] {10.1086/589963}, \href
  {http://adsabs.harvard.edu/abs/2008ApJ...684.1240E} {684, 1240}

\bibitem[\protect\citeauthoryear{{Fall}, {Krumholz}  \& {Matzner}}{{Fall}
  et~al.}{2010}]{fall:2010.sf.eff.vs.surfacedensity}
{Fall} S.~M.,  {Krumholz} M.~R.,   {Matzner} C.~D.,  2010, \mn@doi [\apjl]
  {10.1088/2041-8205/710/2/L142}, \href
  {http://adsabs.harvard.edu/abs/2010ApJ...710L.142F} {710, L142}

\bibitem[\protect\citeauthoryear{{Fatuzzo} \& {Adams}}{{Fatuzzo} \&
  {Adams}}{2008}]{Fatuzzo08}
{Fatuzzo} M.,  {Adams} F.~C.,  2008, \mn@doi [\apj] {10.1086/527469}, \href
  {https://ui.adsabs.harvard.edu/abs/2008ApJ...675.1361F} {675, 1361}

\bibitem[\protect\citeauthoryear{{France} et~al.,}{{France}
  et~al.}{2013}]{2013ApJ...763..149F}
{France} K.,  et~al., 2013, \mn@doi [\apj] {10.1088/0004-637X/763/2/149}, \href
  {https://ui.adsabs.harvard.edu/abs/2013ApJ...763..149F} {763, 149}

\bibitem[\protect\citeauthoryear{{France}, {Arulanantham}, {Fossati}, {Lanza},
  {Loyd}, {Redfield}  \& {Schneider}}{{France}
  et~al.}{2018}]{2018ApJS..239...16F}
{France} K.,  {Arulanantham} N.,  {Fossati} L.,  {Lanza} A.~F.,  {Loyd}
  R.~O.~P.,  {Redfield} S.,   {Schneider} P.~C.,  2018, \mn@doi [\apjs]
  {10.3847/1538-4365/aae1a3}, \href
  {https://ui.adsabs.harvard.edu/abs/2018ApJS..239...16F} {239, 16}

\bibitem[\protect\citeauthoryear{{Gaches} \& {Offner}}{{Gaches} \&
  {Offner}}{2018}]{Gaches18}
{Gaches} B. A.~L.,  {Offner} S. S.~R.,  2018, \mn@doi [\apj]
  {10.3847/1538-4357/aaaae2}, \href
  {https://ui.adsabs.harvard.edu/abs/2018ApJ...854..156G} {854, 156}

\bibitem[\protect\citeauthoryear{{Genzel} et~al.}{{Genzel}
  et~al.}{2010}]{genzel:2010.ks.law}
{Genzel} R.,  et~al., 2010, \mn@doi [\mnras]
  {10.1111/j.1365-2966.2010.16969.x}, \href
  {http://adsabs.harvard.edu/abs/2010MNRAS.407.2091G} {407, 2091}

\bibitem[\protect\citeauthoryear{{Grasha} et~al.,}{{Grasha}
  et~al.}{2019}]{2019MNRAS.483.4707G}
{Grasha} K.,  et~al., 2019, \mn@doi [\mnras] {10.1093/mnras/sty3424}, \href
  {https://ui.adsabs.harvard.edu/abs/2019MNRAS.483.4707G} {483, 4707}

\bibitem[\protect\citeauthoryear{{Grudi{\'c}}, {Hopkins}, {Lee}, {Murray},
  {Faucher-Gigu{\`e}re}  \& {Johnson}}{{Grudi{\'c}}
  et~al.}{2018a}]{grudic:sfe.gmcs.vs.obs}
{Grudi{\'c}} M.~Y.,  {Hopkins} P.~F.,  {Lee} E.~J.,  {Murray} N.,
  {Faucher-Gigu{\`e}re} C.-A.,   {Johnson} L.~C.,  2018a, \mnras, in press,
  arXiv:1809.08348, \href {http://adsabs.harvard.edu/abs/2018arXiv180908348G}
  {}

\bibitem[\protect\citeauthoryear{{Grudi{\'c}}, {Hopkins},
  {Faucher-Gigu{\`e}re}, {Quataert}, {Murray}  \& {Kere{\v s}}}{{Grudi{\'c}}
  et~al.}{2018b}]{grudic:sfe.cluster.form.surface.density}
{Grudi{\'c}} M.~Y.,  {Hopkins} P.~F.,  {Faucher-Gigu{\`e}re} C.-A.,  {Quataert}
  E.,  {Murray} N.,   {Kere{\v s}} D.,  2018b, \mn@doi [\mnras]
  {10.1093/mnras/sty035}, \href
  {http://adsabs.harvard.edu/abs/2018MNRAS.475.3511G} {475, 3511}

\bibitem[\protect\citeauthoryear{{Grudi{\'c}}, {Boylan-Kolchin},
  {Faucher-Gigu{\`e}re}  \& {Hopkins}}{{Grudi{\'c}}
  et~al.}{2019}]{grudic:mond.accel.scale.from.stellar.fb}
{Grudi{\'c}} M.~Y.,  {Boylan-Kolchin} M.,  {Faucher-Gigu{\`e}re} C.-A.,
  {Hopkins} P.~F.,  2019, \mnras, in press, arXiv:1910.06345, \href
  {https://ui.adsabs.harvard.edu/abs/2019arXiv191006345G} {p. arXiv:1910.06345}

\bibitem[\protect\citeauthoryear{{Guszejnov}, {Krumholz}  \&
  {Hopkins}}{{Guszejnov} et~al.}{2016}]{guszejnov.2015:feedback.imf.invariance}
{Guszejnov} D.,  {Krumholz} M.~R.,   {Hopkins} P.~F.,  2016, \mn@doi [\mnras]
  {10.1093/mnras/stw315}, \href
  {http://adsabs.harvard.edu/abs/2016MNRAS.458..673G} {458, 673}

\bibitem[\protect\citeauthoryear{{Guszejnov}, {Hopkins}  \&
  {Grudi{\'c}}}{{Guszejnov} et~al.}{2018}]{guszejnov:universal.scalings}
{Guszejnov} D.,  {Hopkins} P.~F.,   {Grudi{\'c}} M.~Y.,  2018, \mn@doi [\mnras]
  {10.1093/mnras/sty920}, \href
  {http://adsabs.harvard.edu/abs/2018MNRAS.477.5139G} {477, 5139}

\bibitem[\protect\citeauthoryear{{Hansen}, {Klein}, {McKee}  \&
  {Fisher}}{{Hansen} et~al.}{2012}]{hansen:2012.lowmass.sf.radsims}
{Hansen} C.~E.,  {Klein} R.~I.,  {McKee} C.~F.,   {Fisher} R.~T.,  2012,
  \mn@doi [\apj] {10.1088/0004-637X/747/1/22}, \href
  {http://adsabs.harvard.edu/abs/2012arXiv1201.2751H} {747, 22}

\bibitem[\protect\citeauthoryear{{Hartmann}}{{Hartmann}}{2002}]{hartmann:2002.taurus.stellar.clustering}
{Hartmann} L.,  2002, \mn@doi [\apj] {10.1086/342657}, \href
  {http://adsabs.harvard.edu/abs/2002ApJ...578..914H} {578, 914}

\bibitem[\protect\citeauthoryear{{Hennekemper}, {Gouliermis}, {Henning},
  {Brandner}  \& {Dolphin}}{{Hennekemper}
  et~al.}{2008}]{hennekemper:2008.smc.stellar.clustering}
{Hennekemper} E.,  {Gouliermis} D.~A.,  {Henning} T.,  {Brandner} W.,
  {Dolphin} A.~E.,  2008, \mn@doi [\apj] {10.1086/524105}, \href
  {http://adsabs.harvard.edu/abs/2008ApJ...672..914H} {672, 914}

\bibitem[\protect\citeauthoryear{{Heyer}, {Krawczyk}, {Duval}  \&
  {Jackson}}{{Heyer} et~al.}{2009}]{heyer:2009.gmc.trends.w.surface.density}
{Heyer} M.,  {Krawczyk} C.,  {Duval} J.,   {Jackson} J.~M.,  2009, \mn@doi
  [\apj] {10.1088/0004-637X/699/2/1092}, \href
  {http://adsabs.harvard.edu/abs/2009ApJ...699.1092H} {699, 1092}

\bibitem[\protect\citeauthoryear{{Holden}, {Landis}, {Spitzig}  \&
  {Adams}}{{Holden} et~al.}{2011}]{Holden11}
{Holden} L.,  {Landis} E.,  {Spitzig} J.,   {Adams} F.~C.,  2011, \mn@doi
  [\pasp] {10.1086/658081}, \href
  {https://ui.adsabs.harvard.edu/abs/2011PASP..123...14H} {123, 14}

\bibitem[\protect\citeauthoryear{{Hopkins}}{{Hopkins}}{2012}]{hopkins:excursion.ism}
{Hopkins} P.~F.,  2012, \mn@doi [\mnras] {10.1111/j.1365-2966.2012.20730.x},
  \href {http://adsabs.harvard.edu/abs/2012MNRAS.423.2016H} {423, 2016}

\bibitem[\protect\citeauthoryear{{Hopkins}}{{Hopkins}}{2013}]{hopkins:excursion.clustering}
{Hopkins} P.~F.,  2013, \mn@doi [\mnras] {10.1093/mnras/sts147}, \href
  {http://adsabs.harvard.edu/abs/2012arXiv1202.2122H} {428, 1950}

\bibitem[\protect\citeauthoryear{{Hughes} et~al.,}{{Hughes}
  et~al.}{2013}]{2013ApJ...779...46H}
{Hughes} A.,  et~al., 2013, \mn@doi [\apj] {10.1088/0004-637X/779/1/46}, \href
  {https://ui.adsabs.harvard.edu/abs/2013ApJ...779...46H} {779, 46}

\bibitem[\protect\citeauthoryear{{Kennicutt}}{{Kennicutt}}{1998}]{kennicutt98}
{Kennicutt} Jr. R.~C.,  1998, \mn@doi [\apj] {10.1086/305588}, \href
  {http://adsabs.harvard.edu/cgi-bin/nph-bib_query?bibcode=1998ApJ...498..541K&db_key=AST}
  {498, 541}

\bibitem[\protect\citeauthoryear{{Klessen} \& {Burkert}}{{Klessen} \&
  {Burkert}}{2000}]{klessen:2000.cluster.formation}
{Klessen} R.~S.,  {Burkert} A.,  2000, \mn@doi [\apjs] {10.1086/313371}, \href
  {http://adsabs.harvard.edu/abs/2000ApJS..128..287K} {128, 287}

\bibitem[\protect\citeauthoryear{{Kraus} \& {Hillenbrand}}{{Kraus} \&
  {Hillenbrand}}{2008}]{kraus:2008.stellar.clustering}
{Kraus} A.~L.,  {Hillenbrand} L.~A.,  2008, \mn@doi [\apjl] {10.1086/593012},
  \href {http://adsabs.harvard.edu/abs/2008ApJ...686L.111K} {686, L111}

\bibitem[\protect\citeauthoryear{{Kroupa}}{{Kroupa}}{2001}]{kroupa:2001.imf.var}
{Kroupa} P.,  2001, \mn@doi [\mnras] {10.1046/j.1365-8711.2001.04022.x}, \href
  {http://adsabs.harvard.edu/abs/2001MNRAS.322..231K} {322, 231}

\bibitem[\protect\citeauthoryear{{Kryukova}, {Megeath}, {Gutermuth}, {Pipher},
  {Allen}, {Allen}, {Myers}  \& {Muzerolle}}{{Kryukova}
  et~al.}{2012}]{Kryukova12}
{Kryukova} E.,  {Megeath} S.~T.,  {Gutermuth} R.~A.,  {Pipher} J.,  {Allen}
  T.~S.,  {Allen} L.~E.,  {Myers} P.~C.,   {Muzerolle} J.,  2012, \mn@doi [\aj]
  {10.1088/0004-6256/144/2/31}, \href
  {https://ui.adsabs.harvard.edu/abs/2012AJ....144...31K} {144, 31}

\bibitem[\protect\citeauthoryear{{Lee} \& {Chiang}}{{Lee} \&
  {Chiang}}{2017}]{Lee17}
{Lee} E.~J.,  {Chiang} E.,  2017, \mn@doi [\apj] {10.3847/1538-4357/aa6fb3},
  \href {https://ui.adsabs.harvard.edu/abs/2017ApJ...842...40L} {842, 40}

\bibitem[\protect\citeauthoryear{{Lee}, {Miville-Desch{\^e}nes}  \&
  {Murray}}{{Lee} et~al.}{2016}]{lee:mw.cloud.dynamical.sfe}
{Lee} E.~J.,  {Miville-Desch{\^e}nes} M.-A.,   {Murray} N.~W.,  2016, \mn@doi
  [\apj] {10.3847/1538-4357/833/2/229}, \href
  {http://adsabs.harvard.edu/abs/2016ApJ...833..229L} {833, 229}

\bibitem[\protect\citeauthoryear{{Leitherer} et~al.}{{Leitherer}
  et~al.}{1999}]{starburst99}
{Leitherer} C.,  et~al., 1999, \mn@doi [\apjs] {10.1086/313233}, \href
  {http://adsabs.harvard.edu/abs/1999ApJS..123....3L} {123, 3}

\bibitem[\protect\citeauthoryear{{Li}, {Vogelsberger}, {Marinacci}  \&
  {Gnedin}}{{Li} et~al.}{2019}]{2019MNRAS.487..364L}
{Li} H.,  {Vogelsberger} M.,  {Marinacci} F.,   {Gnedin} O.~Y.,  2019, \mn@doi
  [\mnras] {10.1093/mnras/stz1271}, \href
  {https://ui.adsabs.harvard.edu/abs/2019MNRAS.487..364L} {487, 364}

\bibitem[\protect\citeauthoryear{{Marks} \& {Kroupa}}{{Marks} \&
  {Kroupa}}{2012}]{Marks12}
{Marks} M.,  {Kroupa} P.,  2012, \mn@doi [\aap] {10.1051/0004-6361/201118231},
  \href {https://ui.adsabs.harvard.edu/abs/2012A&A...543A...8M} {543, A8}

\bibitem[\protect\citeauthoryear{{Miville-Desch{\^e}nes}, {Duc}, {Marleau},
  {Cuillandre}, {Didelon}, {Gwyn}  \& {Karabal}}{{Miville-Desch{\^e}nes}
  et~al.}{2016}]{2016A&A...593A...4M}
{Miville-Desch{\^e}nes} M.~A.,  {Duc} P.~A.,  {Marleau} F.,  {Cuillandre}
  J.~C.,  {Didelon} P.,  {Gwyn} S.,   {Karabal} E.,  2016, \mn@doi [\aap]
  {10.1051/0004-6361/201628503}, \href
  {https://ui.adsabs.harvard.edu/abs/2016A&A...593A...4M} {593, A4}

\bibitem[\protect\citeauthoryear{{Miville-Desch{\^e}nes}, {Murray}  \&
  {Lee}}{{Miville-Desch{\^e}nes} et~al.}{2017}]{mamd17}
{Miville-Desch{\^e}nes} M.-A.,  {Murray} N.,   {Lee} E.~J.,  2017, \mn@doi
  [\apj] {10.3847/1538-4357/834/1/57}, \href
  {https://ui.adsabs.harvard.edu/abs/2017ApJ...834...57M} {834, 57}

\bibitem[\protect\citeauthoryear{{Moe}, {Kratter}  \& {Badenes}}{{Moe}
  et~al.}{2019}]{Moe19}
{Moe} M.,  {Kratter} K.~M.,   {Badenes} C.,  2019, \mn@doi [\apj]
  {10.3847/1538-4357/ab0d88}, \href
  {https://ui.adsabs.harvard.edu/abs/2019ApJ...875...61M} {875, 61}

\bibitem[\protect\citeauthoryear{{Nakajima}, {Tachihara}, {Hanawa}  \&
  {Nakano}}{{Nakajima} et~al.}{1998}]{nakajima:1998.stellar.clustering}
{Nakajima} Y.,  {Tachihara} K.,  {Hanawa} T.,   {Nakano} M.,  1998, \mn@doi
  [\apj] {10.1086/305493}, \href
  {http://adsabs.harvard.edu/abs/1998ApJ...497..721N} {497, 721}

\bibitem[\protect\citeauthoryear{{Offner}, {Clark}, {Hennebelle}, {Bastian},
  {Bate}, {Hopkins}, {Moraux}  \& {Whitworth}}{{Offner}
  et~al.}{2013}]{offner:2013.imf.review}
{Offner} S.~S.~R.,  {Clark} P.~C.,  {Hennebelle} P.,  {Bastian} N.,  {Bate}
  M.~R.,  {Hopkins} P.~F.,  {Moraux} E.,   {Whitworth} A.~P.,  2013, Protostars
  and Planets VI, University of Arizona Press (2014), eds. H. Beuther, R. S.
  Klessen, C. P. Dullemond, Th. Henning (arXiv:1312.5326), \href
  {http://adsabs.harvard.edu/abs/2013arXiv1312.5326O} {}

\bibitem[\protect\citeauthoryear{{Petigura} et~al.,}{{Petigura}
  et~al.}{2018}]{Petigura18}
{Petigura} E.~A.,  et~al., 2018, \mn@doi [\aj] {10.3847/1538-3881/aaa54c},
  \href {https://ui.adsabs.harvard.edu/abs/2018AJ....155...89P} {155, 89}

\bibitem[\protect\citeauthoryear{{Rice}, {Goodman}, {Bergin}, {Beaumont}  \&
  {Dame}}{{Rice} et~al.}{2016}]{rice:2016.gmc.mw.catalogue}
{Rice} T.~S.,  {Goodman} A.~A.,  {Bergin} E.~A.,  {Beaumont} C.,   {Dame}
  T.~M.,  2016, \mn@doi [\apj] {10.3847/0004-637X/822/1/52}, \href
  {http://adsabs.harvard.edu/abs/2016ApJ...822...52R} {822, 52}

\bibitem[\protect\citeauthoryear{{Rosolowsky}}{{Rosolowsky}}{2005}]{rosolowsky:gmc.mass.spectrum}
{Rosolowsky} E.,  2005, \mn@doi [\pasp] {10.1086/497582}, \href
  {http://adsabs.harvard.edu/abs/2005PASP..117.1403R} {117, 1403}

\bibitem[\protect\citeauthoryear{{Simon}}{{Simon}}{1997}]{simon:1997.stellar.clustering}
{Simon} M.,  1997, \mn@doi [\apjl] {10.1086/310678}, \href
  {http://adsabs.harvard.edu/abs/1997ApJ...482L..81S} {482, L81+}

\bibitem[\protect\citeauthoryear{{Stanke}, {Smith}, {Gredel}  \&
  {Khanzadyan}}{{Stanke} et~al.}{2006}]{stanke:2006.core.mf.clustering}
{Stanke} T.,  {Smith} M.~D.,  {Gredel} R.,   {Khanzadyan} T.,  2006, \mn@doi
  [\aap] {10.1051/0004-6361:20041331}, \href
  {http://adsabs.harvard.edu/abs/2006A%26A...447..609S} {447, 609}

\bibitem[\protect\citeauthoryear{{Thompson}}{{Thompson}}{2013}]{Thompson13}
{Thompson} T.~A.,  2013, \mn@doi [\mnras] {10.1093/mnras/stt102}, \href
  {https://ui.adsabs.harvard.edu/abs/2013MNRAS.431...63T} {431, 63}

\bibitem[\protect\citeauthoryear{{Williams} \& {McKee}}{{Williams} \&
  {McKee}}{1997}]{williams:1997.gmc.prop}
{Williams} J.~P.,  {McKee} C.~F.,  1997, \mn@doi [\apj] {10.1086/303588}, \href
  {http://adsabs.harvard.edu/abs/1997ApJ...476..166W} {476, 166}

\bibitem[\protect\citeauthoryear{{Winter}, {Clarke}, {Rosotti}, {Ih},
  {Facchini}  \& {Haworth}}{{Winter} et~al.}{2018}]{Winter18}
{Winter} A.~J.,  {Clarke} C.~J.,  {Rosotti} G.,  {Ih} J.,  {Facchini} S.,
  {Haworth} T.~J.,  2018, \mn@doi [\mnras] {10.1093/mnras/sty984}, \href
  {https://ui.adsabs.harvard.edu/abs/2018MNRAS.478.2700W} {478, 2700}

\bibitem[\protect\citeauthoryear{{Winter}, {Kruijssen}, {Chevance}, {Keller}
  \& {Longmore}}{{Winter} et~al.}{2020}]{Winter20}
{Winter} A.~J.,  {Kruijssen} J.~M.~D.,  {Chevance} M.,  {Keller} B.~W.,
  {Longmore} S.~N.,  2020, \mn@doi [\mnras] {10.1093/mnras/stz2747}, \href
  {https://ui.adsabs.harvard.edu/abs/2020MNRAS.491..903W} {491, 903}

\bibitem[\protect\citeauthoryear{{Wong} et~al.,}{{Wong}
  et~al.}{2011}]{2011ApJS..197...16W}
{Wong} T.,  et~al., 2011, \mn@doi [\apjs] {10.1088/0067-0049/197/2/16}, \href
  {https://ui.adsabs.harvard.edu/abs/2011ApJS..197...16W} {197, 16}

\bibitem[\protect\citeauthoryear{{Wong} et~al.,}{{Wong}
  et~al.}{2019}]{2019arXiv190511827W}
{Wong} T.,  et~al., 2019, \mnras, in press, arXiv:1905.11827, \href
  {https://ui.adsabs.harvard.edu/abs/2019arXiv190511827W} {p. arXiv:1905.11827}

\bibitem[\protect\citeauthoryear{{Zhang}, {Blake}  \& {Bergin}}{{Zhang}
  et~al.}{2015}]{Zhang15}
{Zhang} K.,  {Blake} G.~A.,   {Bergin} E.~A.,  2015, \mn@doi [\apjl]
  {10.1088/2041-8205/806/1/L7}, \href
  {https://ui.adsabs.harvard.edu/abs/2015ApJ...806L...7Z} {806, L7}

\makeatother
\end{thebibliography}

\end{document}